\documentclass[12pt]{article}
\pdfoutput=1
\usepackage{subfigure}
\usepackage{amssymb,amsmath}
\usepackage{graphicx}
\usepackage{color}
\usepackage[a4paper]{geometry}
\usepackage{placeins}

\makeatletter \renewcommand{\@dotsep}{10000} \makeatother

\mathchardef\mhyphen="2D

\newcommand{\beq}{\begin{equation}}
\newcommand{\eeq}{\end{equation}}
\newcommand{\bea}{\begin{eqnarray}}
\newcommand{\eea}{\end{eqnarray}}

\usepackage[nodisplayskipstretch]{setspace}

\begin{document}

\begin{titlepage}
\pagestyle{empty}

\vspace*{0.2cm}
\begin{center}
{\Large \bf    Light Stops and Fine-Tuning in MSSM
  }\\
\vspace{1cm}
{\bf  Ali \c{C}i\c{c}i$^{a,}$\footnote{E-mail: 501507007@ogr.uludag.edu.tr}, Zerrin K\i rca $^{a,}$\footnote{E-mail: zkirca@uludag.edu.tr} and
Cem Salih $\ddot{\rm U}$n$^{a,}$\footnote{E-mail: cemsalihun@uludag.edu.tr}}
\vspace{0.5cm}

{\it $^a$Department of Physics, Uluda\~{g} University, TR16059 Bursa, Turkey
}

\end{center}

\vspace{0.5cm}
\begin{abstract}
\noindent We discuss the fine-tuning issue within the MSSM framework. Following the idea that the fine-tuning can measure effects of some missing mechanism, we impose non-universal gaugino masses at the GUT scalem and explore the low scale implications. We realize that the fine-tuning parametrized with $\Delta_{EW}$ can be as low as zero. We consider the stop mass with a special importance and focus on the mass scales as $m_{\tilde{t}} \leq 700$ GeV, which are excluded by current experiments when the stop decays into a neutralino along with a top quark or a chargino along with a b quark. We find that the stop mass can be as low as about 200 GeV with $\Delta_{EW} \sim 50$. We find that the solutions in this region can be exluded only up to $60\%$ when stop decays into a neutralino-top quark, and $50\%$ when it decays into a chargino-b quark pair. Setting $65\%$ CL to be potential exclusion and $95\%$ to be pure exclusion limit such solutions will be tested in near future experiments, which are conducted with higher luminosity. In addition to stop, the region with low fine-tuning and light stops predicts masses for the other supersymmetric particles such as $m_{\tilde{b}} \gtrsim 600$ GeV, $m_{\tilde{\tau}} \gtrsim 1$ TeV, $m_{\tilde{\chi}_{1}^{\pm}} \gtrsim 120$ GeV. The details for the mass scales and decay rates are also provided by tables of benchmark points. 
\end{abstract}

\end{titlepage}

%\baselineskip 36pt

%%%%%%%%%%%%%%%%%%%%%%%%%%s
% Main body
%%%%%%%%%%%%%%%%%%%%%%%%%%

\section{Introduction}
\label{sec:Intro}

The Standard Model (SM) of the elementary particles is one of the most successful theory in physics, which has been being tested and confirmed by the strictest experiments for decades. On the other hand, despite the Higgs boson discovery by the ATLAS \cite{Aad:2012tfa} and CMS \cite{Chatrchyan:2013lba} experiments, the SM can only be an effective theory, since it is problematic in stabilizing the Higgs boson mass against the quadratic divergent radiative corrections. Supersymmetry, one of the forefront candidates for physics beyond the SM, can resolve this severe problem by adding superpartners for the SM particles in minimal supersymmetric version of the SM (MSSM). In addition, the tree gauge couplings of the SM can nicely unify at a scale ($\sim 2\times 10^{16}$ GeV), and hence one can build supersymmetric grand unified theories (GUT) to investigate physics at much higher energy scales. Since the Higgs boson mass is free from the quadratic divergences in the MSSM framework, such GUT models can be linked to the low energy scales through the MSSM renormalization group equations (RGEs), which make possible to explore their low scale implications at the current experiments.

Even though the MSSM predictions can be consistent with the current Higgs boson measurements, they have a strong impact in shaping the fundamental parameter space of MSSM. First of all, the MSSM predicts $m_{h} \lesssim M_{Z}$ for the Higgs boson mass at tree level. This inconsistency requires large radiative corrections to be consistent with $m_{h} \sim 125$ GeV. Since the first two family matter particles have negligible couplings to the Higgs boson, the third family particles play a crucial role in radiative contributions to the Higgs boson mass. Moreover, the sbottom and stau, superpartners of bottom quarks and tau lepton respectively, can easily destabilize the Higgs potential \cite{Carena:2012mw}; thus the stability condition on the Higgs potential allows only minor contributions from these sparticles. On the other hand, contributions from the stop, superpartner of top quark, has more freedom without disturbing the Higgs potential stability. After all, the stop sector forms the main source of large radiative corrections to the Higgs boson mass. In order to realize the Higgs boson of mass about 125 GeV, one needs either multi-TeV stop mass, or relatively large soft supersymmetry breaking (SSB) trilinear $A_{t}-$term \cite{Heinemeyer:2011aa}.

Besides the Higgs boson impact, the stop sector can be constrained further by the null results from the direct searches of sparticles at the Large Hadron Collider (LHC). The exclusion on the stop mass depends on the stop's decay channel. If stop is kinematically allowed only to decay into a charm quark and neutralino, then the stop mass bound can be as low as about 230 GeV \cite{TheATLAScollaboration:2013aia}. The constraint becomes much severer when the stop can decay into a bottom quark, a $W-$ boson and a neutralino. In this case the solutions with stop mass lighter than 650 GeV are excluded \cite{Outschoorn:2013pma}. The most strict channel is the one in which the stop decays into a top quark and a neutralino. This channel bounds the stop mass from below at about 750 GeV \cite{LARI:2014jia}. 

In this context, the current results and constraints yield heavy heavy mass spectrum for the SUSY particles, and it brings us back to the naturalness problem. If one characterizes the natural region in SUSY models with $m_{\tilde{t}_{1}}, m_{\tilde{t}_{2}}, m_{\tilde{b}_{1}} \lesssim 500$ GeV \cite{Papucci:2011wy}, it is clearly not possible to fit MSSM consistently in the natural region. Even if the lightest stop mass can be realized as $m_{\tilde{t}_{1}} \lesssim 500$ GeV, the heaviest stop mass eigenstate should be $m_{\tilde{t}_{2}} \gtrsim 1$ TeV to yield a 125 GeV Higgs boson solution \cite{Carena:2011aa}. Such a large splitting between two mass eigenstates of stop indicates a large mixing between the flavor eigenstates, which is proportional to $A_{t}$. Similarly, sbottom is also found heavier than about 1 TeV. 

One proceeds in the naturalness discussion by considering the required fine-tuning in SUSY models, which is discussed in more details in the next section. In this paper, we consider the MSSM framework with non-universal gauginos ($M_{1}\neq M_{2}\neq M_{3}$) and explore the regions with acceptable fine-tuning. Non-universal SSB mass terms for the gauginos can be realized when the gaugino masses are generated with $F-$terms, which are not singlet under the GUT gauge group \cite{Martin:2009ad}. It has been pointed out in \cite{Demir:2014jqa} that if the bilinear Higgs mixing is set to be negative ($\mu < 0$), then the results exhibit more tendency to yield much lower fine-tuning and even light stop solutions, even as light as top quark. However, in the case with $\mu < 0$, the SUSY particles destructively contribute to muon anomalous magnetic moment of muon (muon $g-2$); thus the results for muon $g-2$ are worse than the SM predictions. This drawback can be avoided by setting also $M_{1}, M_{2} < 0$, where $M_{1}$ and $M_{2}$ are the SSB gaugino masses associated with $U(1)_{Y}$ and $SU(2)_{L}$ respectively.

After the physical implications within the fundamental parameters space are investigated, we focus on the solutions with the stop mass lighter than 700 GeV, and discuss the LHC exclusion for these light stop solutions over some benchmark points. The outline of our paper is the following: We first define the parameter to determine the required fine-tuning at the low scale in section \ref{sec:LowFT}. We also discuss the implications and restrictions from the fine-tuning constraint in this section. Section \ref{sec:Scan} describes the data generation and analyses along with the fundamental parameter space and the experimental constraints employed in our analyzes. Then, we discuss our results for the fine-tuning with highlighting the light stop solutions ($m_{\tilde{t}_{1}} \lesssim 700$ GeV) in Section \ref{sec:Results}. After discussing the impact of the fine-tuning and light stop solutions, we also present the mass spectrum for the other sparticles in Section \ref{sec:Spec}. In Section \ref{sec:LHC} we analyze if the LHC can detect such light stop solutions over some benchmark points. Finally we conclude in Section \ref{sec:conc}.

\section{Low Scale Fine-Tuning Measurement}
\label{sec:LowFT}

Compared to the SM, the Higgs sector is more complicated in the MSSM, since there is two Higgs doublets, which both develop non zero vacuum expectation values (VEVs). Also, it has been shown a long time ago that the SUSY has to be broken to realize the correct EW breaking scale ($\sim 100$ GeV), since the minimization of the Higgs potential requires $m_{H_{u}}\neq m_{H_{d}}$ \cite{Martin:1997ns}. As discussed in the previous section, the fundamental parameter space of MSSM needs to be fine-tuned, and it can be analyzed by considering the $Z-$boson mass with the following equation

\begin{equation}
\frac{1}{2} M_{Z}^{2} = -\mu^{2}+\frac{(m^{2}_{H_{d}}+\Sigma_{d})-(m^{2}_{H_{u}}+\Sigma_{u})\tan^{2}\beta}{\tan^{2}\beta-1}~,
\label{zmass}
\end{equation}
where $\Sigma_{d,u}$ denote the radiative contributions to the SSB Higgs boson masses $m_{H_{d,u}}$. The left hand side of Eq.(\ref{zmass}) is precisely determined by the experiments, while the right hand side is involved with the fundamental parameters of MSSM, whose values can lie in a wide range. Thence, there needs to be significant cancellations among the parameters in the right hand side to yield consistent $M_{Z}$. Since the terms with $m_{H_{d}}$ (and $\Sigma_{d}$) are suppressed by $\tan\beta$, the cancellations happen mainly among the terms with $\mu$ and $m_{H_{u}}$, and the correct EW breaking scale requires $\mu \approx m_{H_{u}}$ over most of the fundamental parameter space. The required amount of fine-tuning can be quantified with $\Delta_{EW}$, which is defined based on Eq.(\ref{zmass}) as 

\begin{equation}
  \Delta_{EW}\equiv {\rm Max}(C_{i})/(M_{Z}^{2}/2)~,\hspace{0.3cm}{\rm where}\hspace{0.3cm} C_{i}=\left\lbrace \begin{array}{ll} C_{H_{d}} = & \mid m_{H_{d}}^{2}/(\tan^{2}\beta -1)\mid \\ & \\
   C_{H_{u}} = & \mid m_{H_{u}}^{2}\tan^{2}\beta/(\tan^{2}\beta -1)\mid \\ & \\
   C_{\mu} = & \mid -\mu^{2} \mid~,
  \end{array}\right.
  \label{FT}
\end{equation}  
here we have assumed that the radiative corrections $\Sigma_{d,u}$ are included in $m_{H_{d,u}}$. In contrast to characterizing the natural region, the amount of fine-tuning does not depend on the sparticle masses directly. However, the sparticle spectrum and mixings among them are still important, since they take part in radiative corrections to $m_{H_{d,u}}$. 

If it is possible to realize low $\mu^{2}$ values over the fundamental parameter space, the fine-tuning can be found in an acceptable range regardless to the sparticle mass spectrum. However, the effects from the sparticle masses are encoded in the radiative corrections. $\Sigma_{d}$ is evolved with the sbottom and stau masses, which contribute to $m_{H_{d}}$ at the loop level. Since this term is suppressed by $\tan\beta$, the effects from the sbottom and stau masses in the fine-tuning are minor. On the other hand, $\Sigma_{u}$, which arises from the stop sector, does not exhibit a suppression by $\tan\beta$.  Large stop masses or large mixings between left and right handed stops can significantly contribute to the radiative corrections which result in large $m_{H_{u}}$, and thus large $\mu-$term. Considering the severe experimental exclusion limits on stops, discussed in the previous section, it is obvious that the parameter space, allowed by the experiments, needs to be largely fine-tuned . Even if one restricts the lightest stop masses to be at a few hundred GeV, then a large mixing between stops is required by the Higgs boson mass. Such a large mixing results in very large radiative corrections, and hence, raises the required fine-tuning significantly \cite{Demir:2014jqa}. This discussion can be concluded that the SUSY models need large fine-tuning when the sparticle and the gaugino masses are set universal at the GUT scale. 

If one relaxes the exclusion limits from the LHC, mentioned above, and allows the solutions with light stop, the required fine-tuning can potentially be improved at the low scale. However, the requirement to yield the Higgs boson of mass about 125 GeV also puts a severe constraint on the stop masses as discussed in the previous section. The Higgs boson mass within the MSSM can be written as 

\begin{equation}
m_{h}\approx M_{Z}\cos\beta+\frac{3m_{t}^{4}}{4\pi^{2}v^{2}}\left(\log \frac{M_{S}^{2}}{m_{t}^{2}} + \frac{X_{t}}{M_{S}^{2}}-\frac{X_{t}^{4}}{12 M_{S}^{4}} \right)-\frac{y_{b}^{4}\mu^{4}v^{2}}{16\pi^{2}M_{S}^{4}}
\label{hmass}
\end{equation}
where $m_{t}$ is the top quark mass, while $M_{S}\equiv \sqrt{m_{\tilde{t}_{L}}m_{\tilde{t}_{R}}}$ is the average stop mass. $M_{S}$ is also the energy scale at which the supersymmetric particles decouple from the SM. The mixing in the stop sector is encoded in $X_{t}$ as $X_{t}=A_{t}-\mu\cot\beta$, where $A_{t}$ stands for this mixing. The first term in Eq.(\ref{hmass}) is the tree-level mass of the Higgs, and it can only be about 90 GeV at most. Thus, it needs significant loop corrections to realize the Higgs boson of mass about 125 GeV. Such large corrections can be obtained with a large mass splitting between the stop and top quarks ($M_{S}\gg m_{t}$). Another way to raise the loop corrections is to implement large mixing in the stop sector. We should note that here $A_{t}\lesssim 3 M_{S}$ should be satisfied not to break color and/or charge conservation at minima of the scalar potential \cite{Ellwanger:1999bv}. Hence, in the case of large mixing, sparticles cannot be lighter than certain mass scales.

The last term in Eq.(\ref{hmass}) represents loop contributions from the bottom sector, but this term is relevant only for large $\tan\beta$. Consequently, the only dominant source for large loop corrections to the Higgs boson mass is the stop sector, which requires the stop  to be heavier even if the mixing in this sector is large.  This situation can be drastically different if MSSM is extended with new particles and/or new symmetries \cite{Gogoladze:2012jp,Gogoladze:2014vea,Elsayed:2011de,Khalil:2015wua,Li:2015dil,Hicyilmaz:2016kty} which contribute to the Higgs boson mass as significantly as the stop. In this context, the minimal supersymmetric models may not cover the full picture of physics. The mechanisms, which are not included in the minimal models, can effect the low scale phenomenology. In this sense, the fine-tuning requirement can emerge because of some missing mechanisms, and its amount can be interpreted the effectiveness of these missing mechanisms, and also indicates the amount of deviation from the minimality. The effects from missing mechanisms can be analyzed also within the MSSM framework by implementing non-universalities in gaugino and scalar sectors \cite{Gogoladze:2012yf,Gogoladze:2013wva,Calibbi:2016qwt,Gogoladze:2016grr}

In our work, we analyze the effects of possible missing mechanisms within the MSSM framework by imposing non-universality in the gaugino sector. While we focus on the regions with low fine-tuning, we also highlight the stop  masses less than 700 GeV, and discuss if such solutions can still survive under the severe experimental constraints.

\section{Scanning Procedure and Experimental Constraints}
\label{sec:Scan}

We have employed SPheno 3.3.8 package \cite{Porod:2003um,Porod2} obtained with SARAH 4.5.8 \cite{Staub:2008uz,Staub2}. In this package, the weak scale values of the gauge and Yukawa couplings presence in MSSM are evolved to the unification scale $M_{{\rm GUT}}$ via the renormalization group equations (RGEs). $M_{{\rm GUT}}$ is determined by the requirement of the gauge coupling unification through their RGE evolutions. Note that we do not strictly enforce the unification condition $g_1 = g_2 = g_3$ at $M_{{\rm GUT}}$ since a few percent deviation from the unification can be assigned to unknown GUT-scale threshold corrections \cite{Hisano:1992jj,GUTth}. With the boundary conditions given at $M_{{\rm GUT}}$, all the SSB parameters along with the gauge and Yukawa couplings are evolved back to the weak scale.

We have performed random scans over the following parameter space

\begin{equation}
%\setstretch{1.5}
\begin{array}{ccc}
0 \leq & m_{0} & \leq 10~{\rm TeV} \\
-10 \leq & M_{1} & \leq 0~{\rm TeV} \\
-10 \leq & M_{2} & \leq 0~{\rm TeV} \\
0 \leq & M_{3} & \leq 10~{\rm TeV} \\
-3  \leq & A_{0}/m_{0} &\leq 3  \\
 2  \leq & \tan\beta & \leq 60
\end{array}
\label{paramSpace}
\end{equation}
\begin{equation*}
\mu < 0~,\hspace{0.3cm} m_{t}=173.3~{\rm GeV}
\end{equation*}

where $m_0$ is the universal SSB mass term for the matter scalars and Higgs fields. $M_3$, $M_2$ and $M_1$ are SSB mass terms for the gauginos associated with the  SU(3), SU(2) and U(1) symmetry groups respectively. $A_0$ is SSB trilinear coupling, and $\tan\beta$ is ratio of VEVs of the MSSM Higgs doublets. In the constrained MSSM (CMSSM) with non-universal gauginos all matter scalars have the same mass and gaugino masses can be chosen different each other at the GUT scale. The radiative EW breaking (REWSB) condition determines the value of $\mu-term$ but not its sign; thus, its sign is one of the free parameters, and we set it negative in our scans. In addition, we have used central value of top quark mass as $m_t=173.3$ GeV \cite{Group:2009ad}. Note that the sparticle spectrum is not too sensitive in one or two sigma variation in the top quark mass \cite{Gogoladze:2011db}, but it can shift the Higgs boson mass by  1-2 GeV \cite{Gogoladze:2011aa,Ajaib:2013zha}.

The REWSB condition provides a strict theoretical constraint \cite{Ibanez:Ross,REWSB2,REWSB3,REWSB4,REWSB5} over the fundamental parameter space given in Eq.(\ref{paramSpace}). Another important constraint comes from the relic abundance of charged supersymmetric particles \cite{Nakamura:2010zzi}. This constraint excludes the regions which yield charged particles such as stop and stau being the lightest supersymmetric particle (LSP). In this context, we accept only the solutions which satisfy the REWSB condition and yield neutralino LSP. When one requires the solutions to yield one of the neutralinos to be LSP, it si also suitable that the LSP can be promoted as a candidate for dark matter. In this case, the relic abundance of LSP should also be consistent with the current results from the WMAP \cite{Hinshaw:2012aka} and Planck \cite{Ade:2015xua} satellites. However, even if a solution does not satisfy the dark matter observations, it can still survive in conjunction with other form(s) of the  dark matter formation \cite{Baer:2012by}. In this context, we do not require the solutions to satisfy the WMAP or Planck results on the relic abundance of LSP neutralino in our analyses.

In scanning the parameter space we use our interface, which employs Metropolis-Hasting algorithm described in  \cite{Belanger:2009ti,SekmenMH}. After collecting the data, we successively apply the mass bounds on all sparticles \cite{Agashe:2014kda} and the constraints from the rare B-decays ($B_s \rightarrow \mu^+ \mu^-$ \cite{Aaij:2012nna}, $B_s \rightarrow X_s \gamma$ \cite{Amhis:2012bh} and $B_u \rightarrow \tau \nu_\tau $ \cite{Asner:2010qj}). The experimental constraints can be listed as follows:
\begin{equation}
%\setstretch{2.0}
\begin{array}{l}
123 \leq m_{h} \leq 127~{\rm GeV}\\
 m_{\tilde{g}}  \geq 1000~{\rm GeV}\\
 0.8\times 10^{-9} \leq BR (B_s \rightarrow {\mu}^{+} {\mu}^{-}) \leq 6.2 \times 10^{-9}~ (2\sigma) \\
2.9\times 10^{-4} \leq BR (b \rightarrow s {\gamma})\leq 3.87\times 10^{-4}~ (2\sigma)\\
0.15 \leq \dfrac{BR (B_u \rightarrow {\nu}_{\tau} {\tau})_{MSSM}}{BR (B_u \rightarrow {\nu}_{\tau} {\tau})_{SM}}\leq 2.41 ~ (3\sigma)
\end{array}
\label{constraints}
\end{equation}

Note that the mass bound on the gluino listed above is relaxed in compared to the current bounds ($m_{\tilde{g}}\geq 1.9$ TeV \cite{Sirunyan:2017cwe}). The aim in relaxing the mass bound on gluino is to see how light stop can be obtained and how much improvement can be realized in fine-tuning in over all results. When we proceed our analyses over some benchmark points, the current bound on the gluino mass will be taken into account.

One of the strongest constraints are comes from rare B-meson decay into a muon pair. The supersymmetric contributions to the $BR(B_s \rightarrow {\mu}^{+} {\mu}^{-})$ are severely constrained, since the SM's predictions almost overlap with its experimental measurements. Supersymmetric contributions to this process are proportional to $(\tan\beta)^6/m^4_A$. Therefore, it has a strong impact on the regions with large $\tan\beta$ that CP-odd Higgs boson has to be heavy enough ($m_{A}\sim $ TeV) to suppress the supersymmetric contribution. Finally, we also require the solutions to do no worse than the SM in regard of the muon $g-2$ by requiring $\Delta a_{\mu} \geq 0$.

\section{Fine-Tuning and Sparticle Mass Spectrum in MSSM}
\label{sec:Results}

\begin{figure}[ht!]
\centering
\subfigure{\includegraphics[scale=0.8]{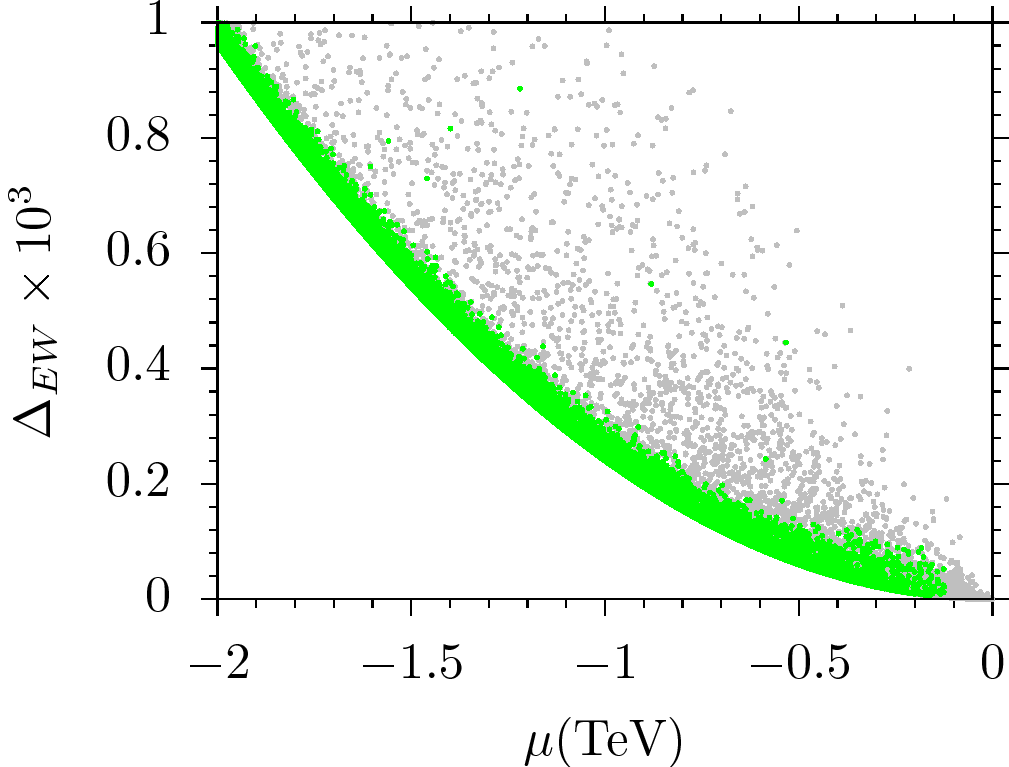}}
\subfigure{\hspace{0.3cm}\includegraphics[scale=0.8]{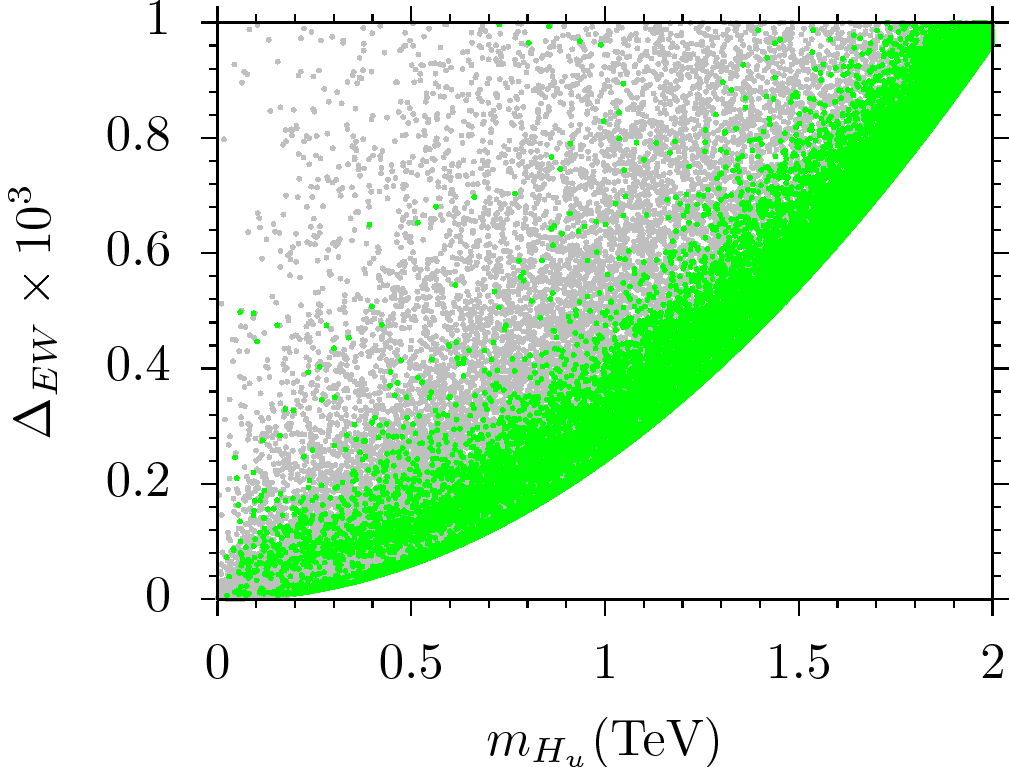}}\\
\subfigure{\includegraphics[scale=0.8]{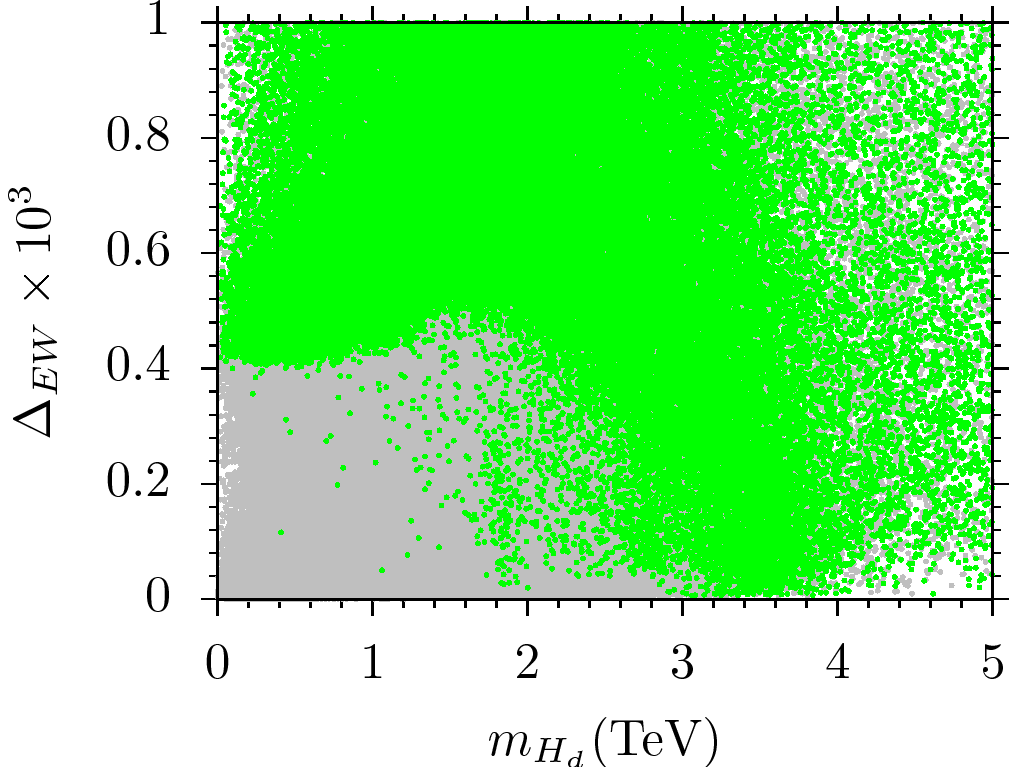}}
\subfigure{\hspace{0.3cm}\includegraphics[scale=0.8]{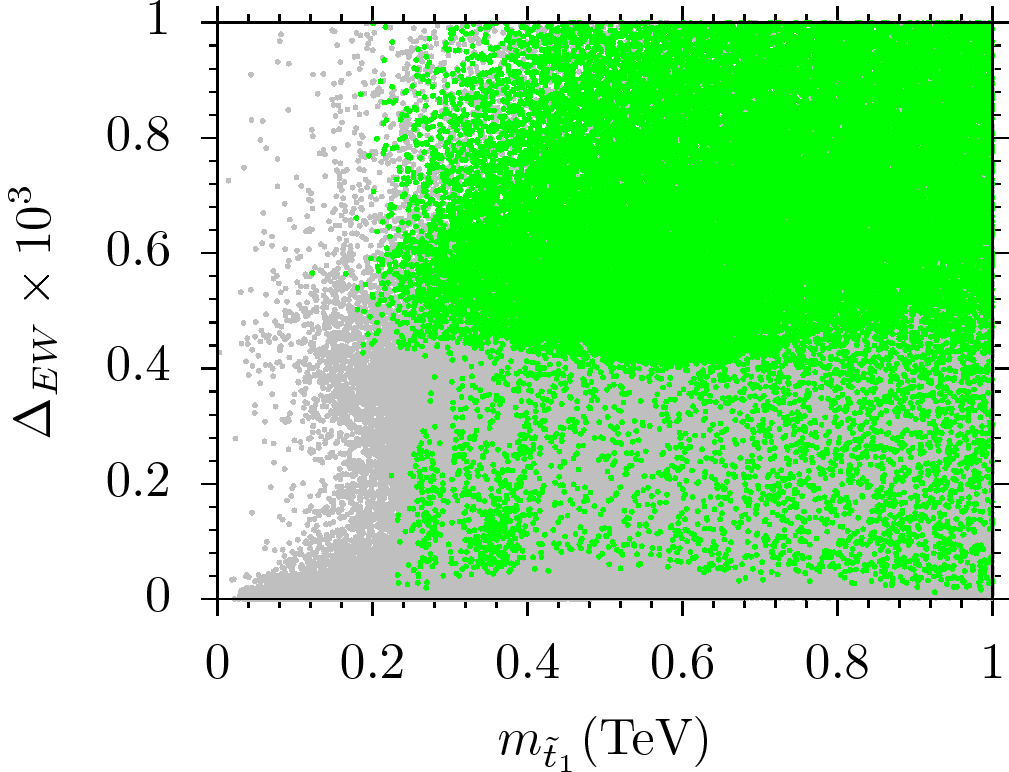}}
\caption{Plots in $\Delta_{EW}-\mu$, $\Delta_{EW}-m_{H_{u}}$, $\Delta_{EW}-m_{H_{d}}$, and $\Delta_{EW}-m_{\tilde{t}}$ planes. All points are consistent with REWSB and neutralino LSP. Gray points are excluded by the current experimental bounds, while the green points are allowed.}
\label{fig1}
\end{figure}

\begin{figure}[t!]
\centering
\subfigure{\includegraphics[scale=0.8]{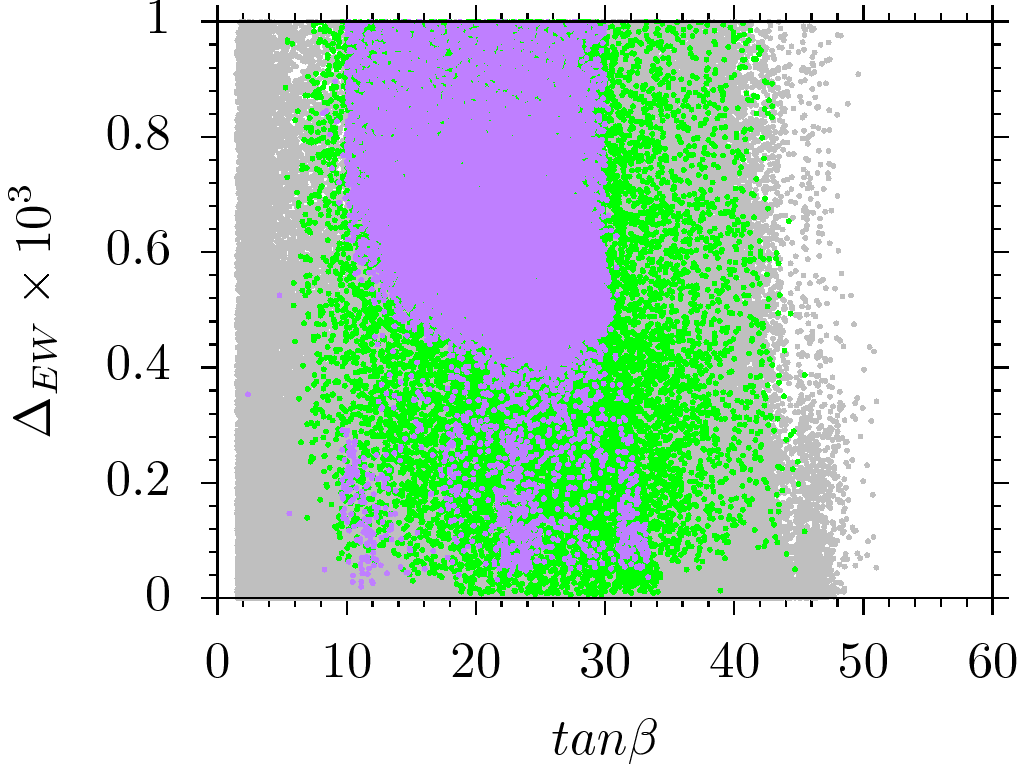}}
\subfigure{\hspace{0.3cm}\includegraphics[scale=0.8]{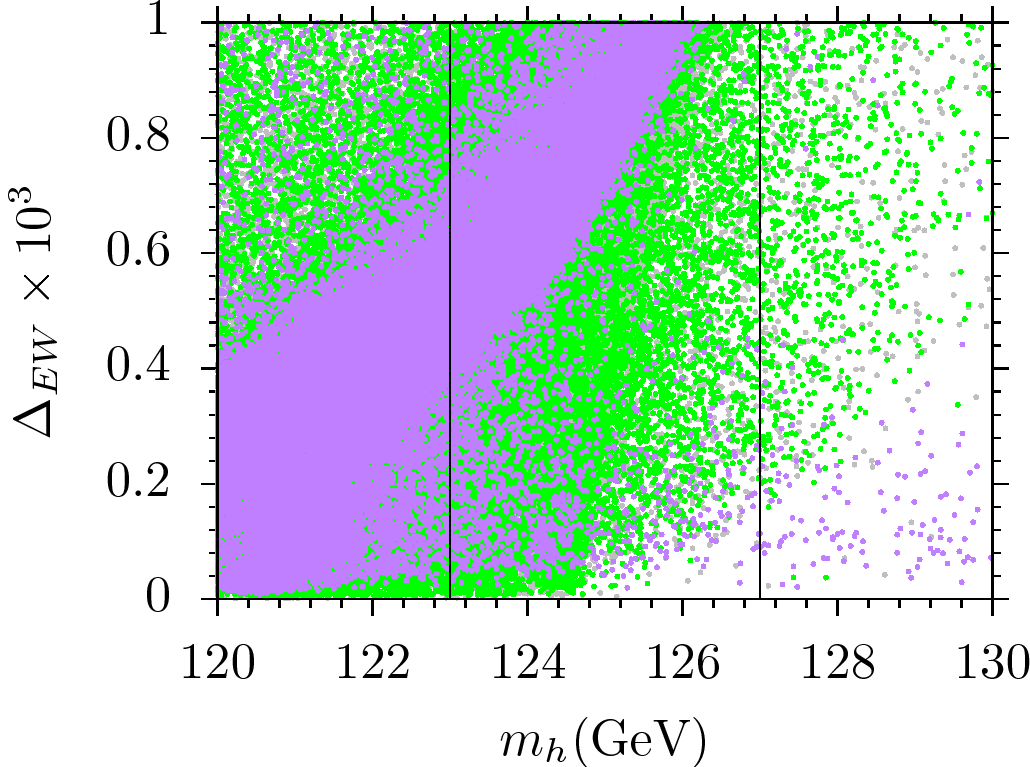}}
\caption{Plots in $\Delta_{EW}-\tan\beta$ and $\Delta_{EW}-m_{h}$ planes. The color coding is the same as Figure \ref{fig1}. In addition, the purple points form a subset of green and they represent the solutions with $m_{\tilde{t}_{1}}\leq 700$ GeV. We do not apply the Higgs mass bound in the $\Delta_{EW}-m_{h}$ plane, since it is represented in one axis. We use rather vertical lines which shows the experimental bounds on the Higgs boson mass.}
\label{fig2}
\end{figure}
\begin{figure}[b!]
\centering
\subfigure{\includegraphics[scale=0.8]{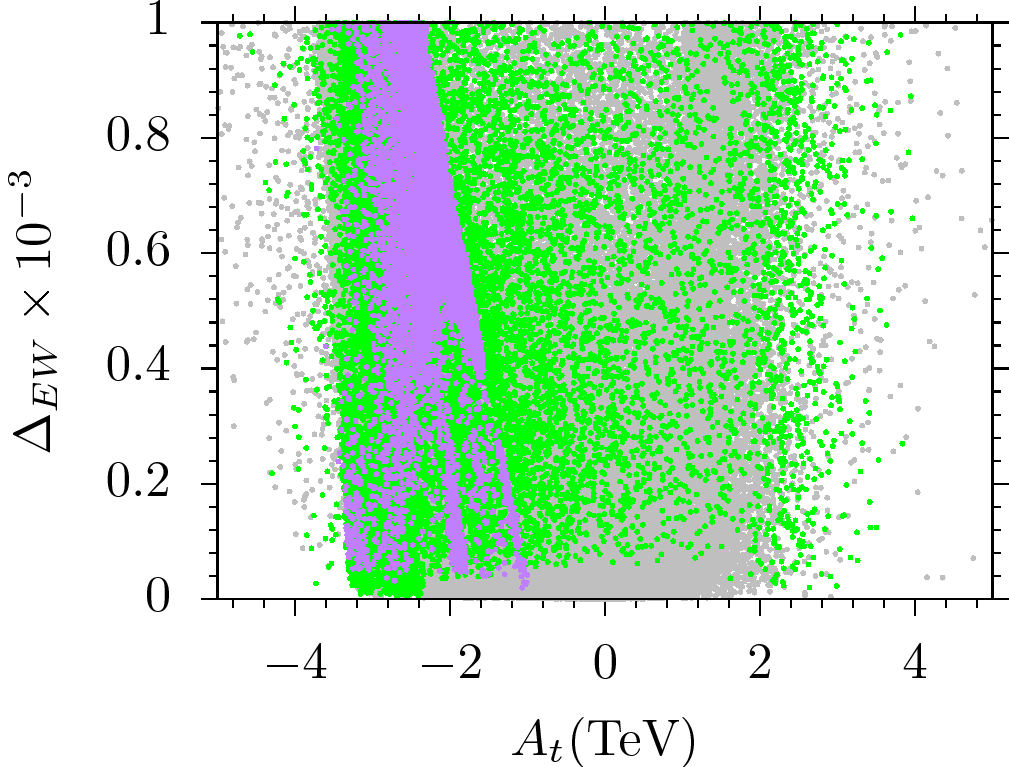}}
\subfigure{\hspace{0.3cm}\includegraphics[scale=0.8]{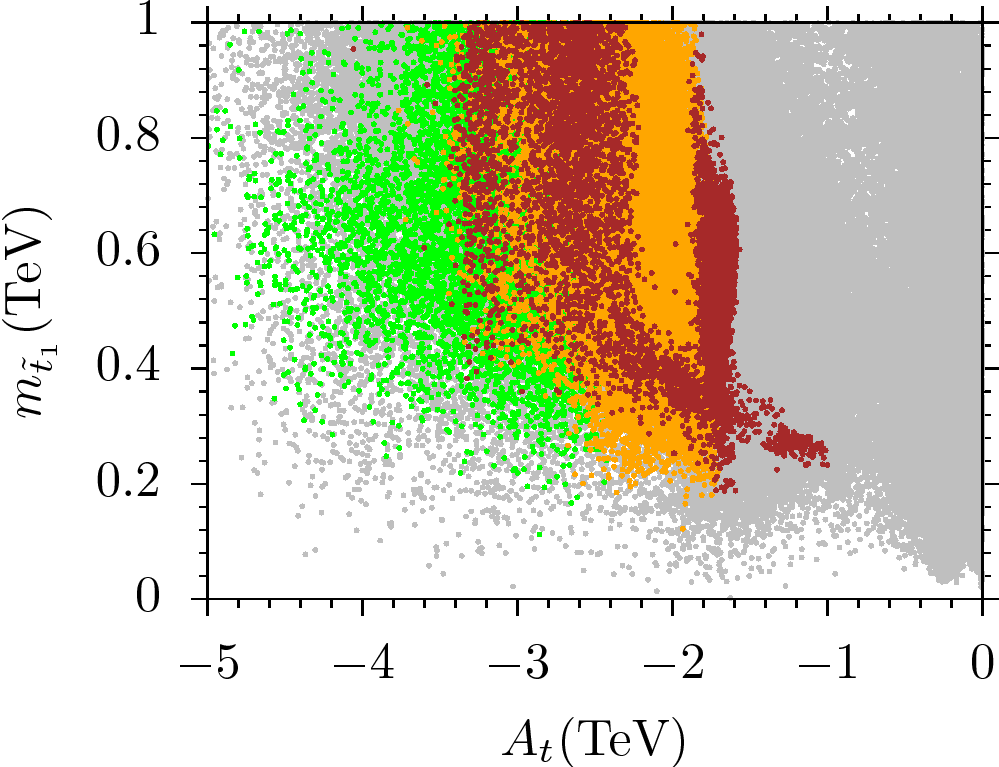}}
\caption{Plots in $\Delta_{EW}-A_{t}$ and $m_{\tilde{t}_{1}}-A_{t}$ planes. The color coding in the left panel is the same as Figure \ref{fig1}. While the meaning of gray and green are the same in the right panel, the orange points represents the solutions with $\Delta_{EW}\leq 10^{3}$, and the brown points form a subset of orange with $\Delta_{EW}\leq 500$. The condition $m_{\tilde{t}_{1}} \leq 700$ GeV is not applied in the right panel, since the stop mass is represented directly in one axis.}
\label{fig3}
\end{figure}

In this section, we present our results for the fine-tuning and the stop mass and highlight if there is any correlation between them. The acceptable fine-tuning amount can be applied conventionally as $\Delta_{EW}\leq 10^{3}$. Figure \ref{fig1} represents our results with plots in $\Delta_{EW}-\mu$, $\Delta_{EW}-m_{H_{u}}$, $\Delta_{EW}-m_{H_{d}}$, and $\Delta_{EW}-m_{\tilde{t}}$ planes. All points are consistent with REWSB and neutralino LSP. Gray points are excluded by the current experimental bounds, while the green points are allowed.The $\Delta_{EW}-\mu$ plane reveals a strong correlation between the fine-tuning and the $\mu-$term, which is seen as a tight parabolic curve. According to these results, the required amount of fine-tuning is beyond the acceptable range when $|\mu| \gtrsim 2$ TeV. A similar correlation can be also realized between $\Delta_{EW}$ and $m_{H_{u}}$, despite not being as strict as that for the $\mu-$term. The results in the $\Delta_{EW}-m_{H_{u}}$ plane is the impact of the correct EW symmetry breaking scale condition which requires $\mu \approx m_{H_{u}}$. The $\Delta_{EW}-m_{H_{d}}$ plane does not show any correlation between $\Delta_{EW}$ and $m_{H_{d}}$, as discussed before that $m_{H_{d}}$ is not very strong in calculating the fine-tuning. Surprisingly, the fine-tuning results do not exhibit a strong correlation with the stop mass as seen from the $\Delta_{EW}-m_{\tilde{t}}$ plane, and it is possible to realize the stop as light as about 200 GeV with very low fine-tuning measures ($\sim 0$). 

Figure \ref{fig2} displays the results with plots in the $\Delta_{EW}-\tan\beta$ and $\Delta_{EW}-m_{h}$ planes. The color coding is the same as Figure \ref{fig1}. In addition, the purple points form a subset of green and they represent the solutions with $m_{\tilde{t}_{1}}\leq 700$ GeV. We do not apply the Higgs mass bound in the $\Delta_{EW}-m_{h}$ plane, since it is represented in one axis. We use rather vertical lines which show the experimental bounds on the Higgs boson mass. The $\Delta_{EW}-\tan\beta$ plane exhibits a restriction in $\tan\beta$ range that this parameter cannot take a value greater than $50$. On the other hand, this restriction on this parameter does not arise from the fine-tuning condition, it is rather related to the REWSB condition. In the allowed range it is possible to obtain low fine-tuning for any value of $\tan\beta$. If one applies another condition on the stop mass such that $m_{\tilde{t}_{1}}\leq 700$ GeV, the solutions restrict the parameter space as $ 10 \lesssim \tan\beta \lesssim 30$ (purple).  We also present the correlation between the fine-tuning and the Higgs boson mass in the $\Delta_{EW}-m_{h}$. The results reflect the fact that if the Higgs boson was found much lighter than its current experimental range, the minimal SUSY models could not suffer from large fine-tuning issues. The $\Delta_{EW}-m_{h}$ plane shows that $\Delta_{EW}$ can be as low as about zero, which means no fine-tuning is required, when the Higgs boson mass is about 120 GeV. However these regions have been already excluded. The experimental results allow only the region between two vertical lines in the plane. One can easily see that the fine-tuning can go worse as the Higgs boson mass increases.

Despite the increasing Higgs boson mass worsens the fine-tuning, there is also another branch through which the fine-tuning remains almost constant ($\Delta_{EW}\sim 100-200$), while the Higgs boson mass increases. The solutions in this branch can be understood by considering the the mixing in the stop sector, which is proportional to $A_{t}$. We discuss this effect with plots in $\Delta_{EW}-A_{t}$ and $m_{\tilde{t}_{1}}-A_{t}$ planes of Figure \ref{fig3}. The color coding in the left panel is the same as Figure \ref{fig1}. While the meaning of gray and green are the same in the right panel, the orange points represents the solutions with $\Delta_{EW}\leq 10^{3}$, and the brown points form a subset of orange with $\Delta_{EW}\leq 500$. The condition $m_{\tilde{t}_{1}} \leq 700$ GeV is not applied in the right panel, since the stop mass is represented directly in one axis. The $\Delta_{EW}-A_{t}$ plane shows that the solutions with acceptable fine-tuning and $m_{\tilde{t}_{1}}\leq 700$ GeV (shown in purple) require $A_{t}$ from about $1$ to $4$ TeV in the negative region. The negative sign of $A_{t}$ reverses the effect of stop mixings in the fine-tuning calculations. In this context, the positive values of $A_{t}$, even in the range $1-4$ TeV causes highly fine-tuned solutions, and hence one cannot found any purple point in the positive $A_{t}$ region. The branch with constant $\Delta_{EW}$ mentioned above can also be seen in the $m_{\tilde{t}_{1}}-A_{t}$ plane. In one branch, $A_{t}$ and the stop mass change together and they yield an increase in the fine-tuning as well. In the second branch, $A_{t}$ remains constant at about $-1.8$ TeV, while the stop mass increases up to about a TeV. The constant $A_{t}$ and fine-tuning with increasing stop mass  in this branch can be concluded that stop mass is not a strong parameter in the fine-tuning, while $A_{t}$ is determining the behavior of the solutions in respect of the fine-tuning.

\section{Sparticle Mass Spectrum}
\label{sec:Spec}

\begin{figure}[b!]
\centering
\subfigure{\includegraphics[scale=0.8]{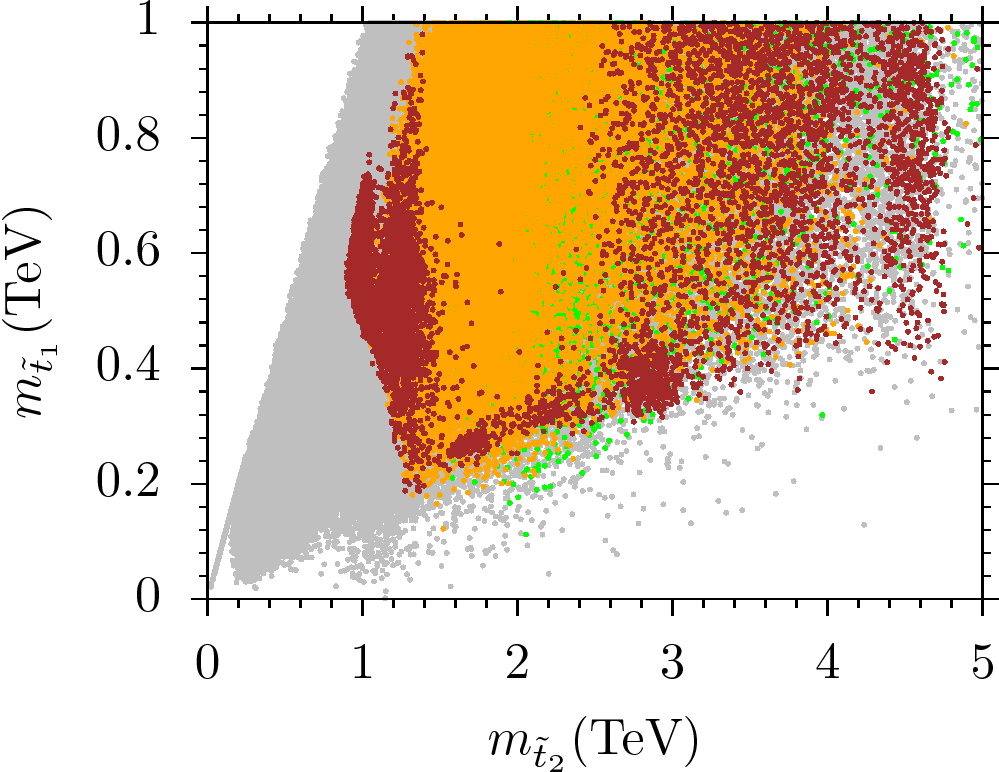}}
\subfigure{\includegraphics[scale=0.8]{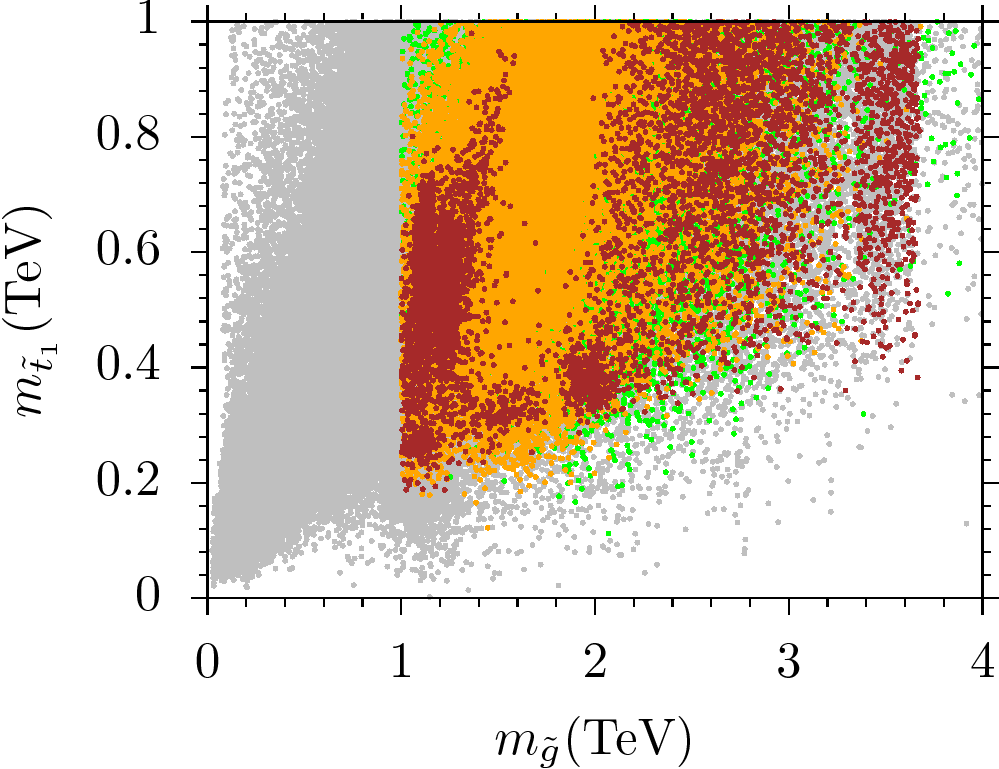}}\\
\subfigure{\includegraphics[scale=0.8]{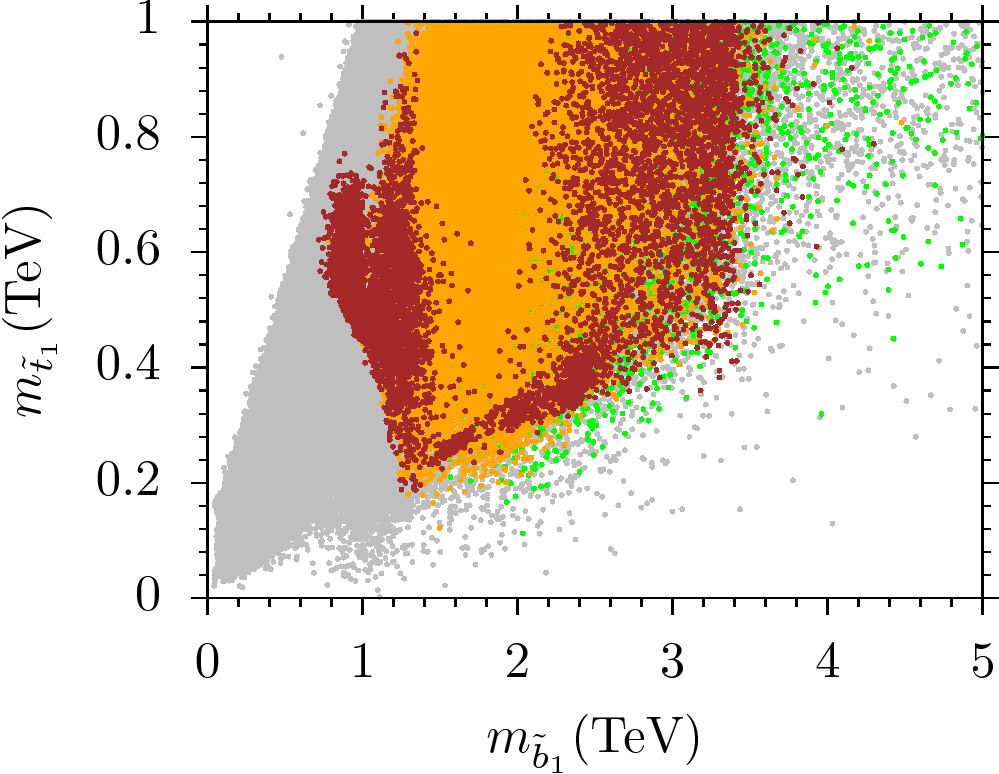}}
\subfigure{\includegraphics[scale=0.8]{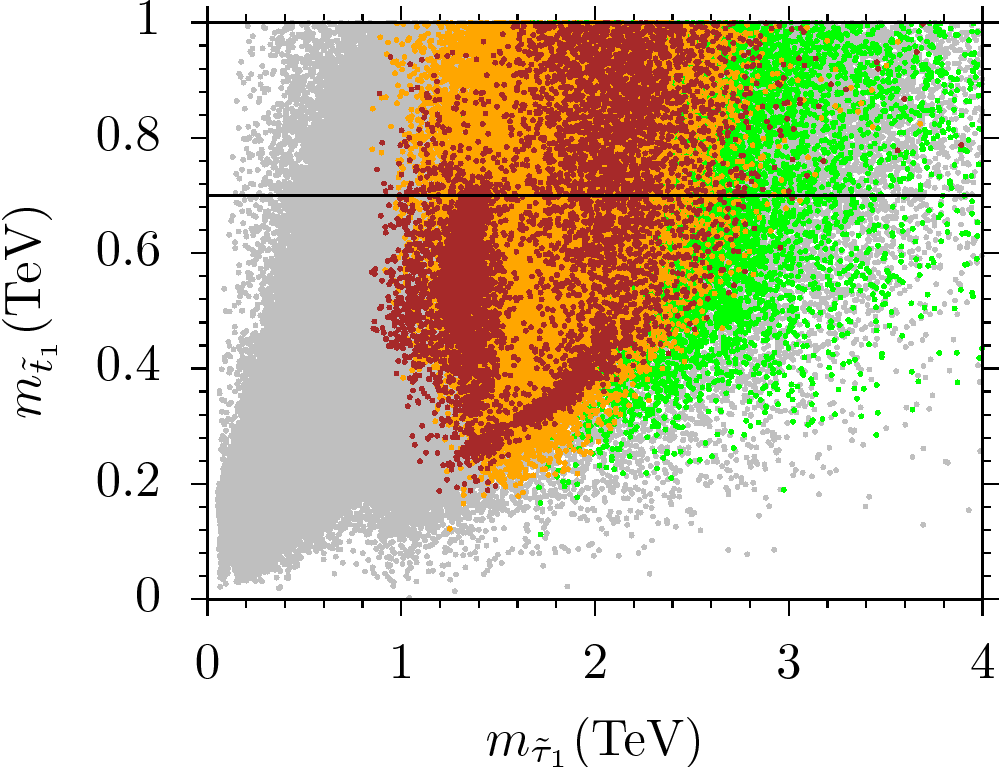}}
\caption{The sparticle masses in $m_{\tilde{t}_{1}}-m_{\tilde{t}_{2}}$, $m_{\tilde{t}_{1}}-m_{\tilde{g}}$, $m_{\tilde{t}_{1}}-m_{\tilde{b}_{1}}$, and $m_{\tilde{t}_{1}}-m_{\tilde{\tau}_{1}}$. The color coding is the same as the right panel of Figure \ref{fig3}.}
\label{fig4}
\end{figure}

In this section, we consider the mass spectrum of the supersymmetric particles in addition to the stop, since they are also of special importance in exploring the low energy implications of MSSM. Figure \ref{fig4} represents masses of stop, gluino, sbottom and stau with plots in the $m_{\tilde{t}_{1}}-m_{\tilde{t}_{2}}$, $m_{\tilde{t}_{1}}-m_{\tilde{g}}$, $m_{\tilde{t}_{1}}-m_{\tilde{b}_{1}}$, and $m_{\tilde{t}_{1}}-m_{\tilde{\tau}_{1}}$. The color coding is the same as the right panel of Figure \ref{fig3}. According to the results represented in the $m_{\tilde{t}_{1}}-m_{\tilde{t}_{2}}$ plane, the second stop cannot be lighter than about a TeV, although the lightest stop can be as light as about 200 GeV. When one of the stops is light, the Higgs boson mass constraint pushes the second stop mass up to the TeV scale or above, which also requires a large mixing in the stop sector. In addition to the mixing, gluino can also lead to heavy stop, since it contributes radiatively to the stop mass. The $m_{\tilde{t}_{1}}-m_{\tilde{g}}$ shows that the stop can be as light as 200 GeV when $m_{\tilde{g}} \sim 1$ TeV. The increase in the stop mass with increasing gluino mass can be seen from the results. However, heavier gluino mass can provide only a slight increase, and it is still possible to realize $m_{\tilde{t}_{1}} \gtrsim 250$ GeV, when $m_{\tilde{g}} \gtrsim 1.9$ TeV. Similarly sbottom and stau cannot be lighter than about 1 TeV, when $m_{\tilde{t}_{1}}\lesssim 400$ GeV as seen from the bottom panels of Figure \ref{fig4}. However, it is possible to realize nearly degenerate stop and sbottom when $m_{\tilde{t}_{1}}\simeq m_{\tilde{b}_{1}} \approx 600$ GeV.

\begin{figure}[h!]
\centering
\subfigure{\includegraphics[scale=0.8]{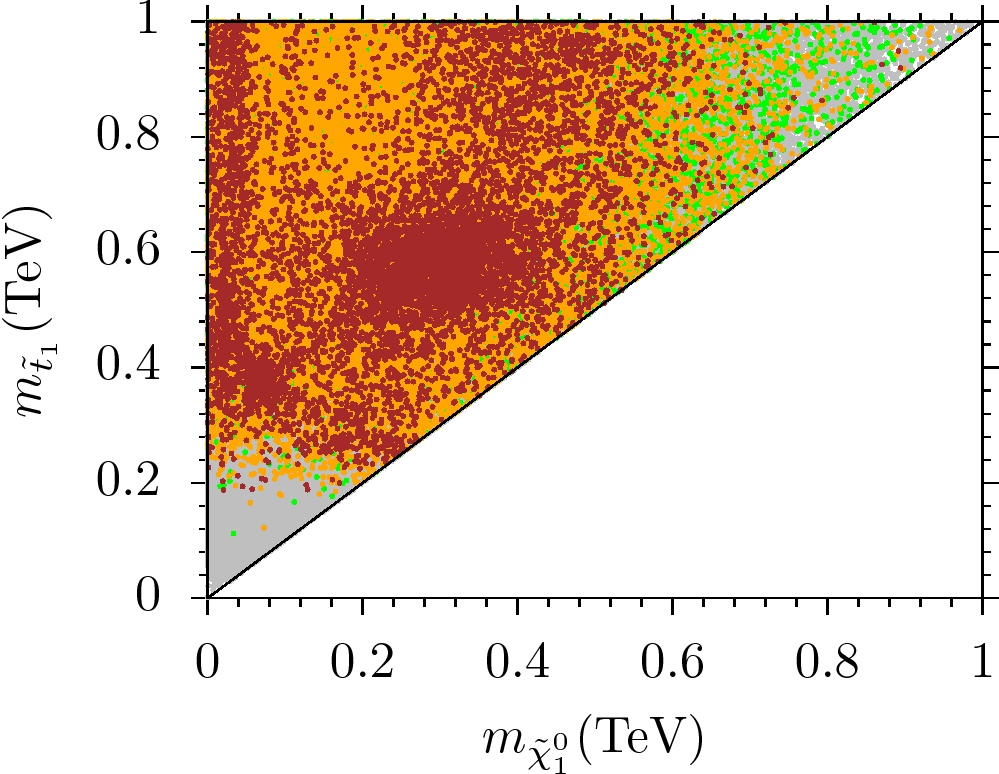}}
\subfigure{\includegraphics[scale=0.8]{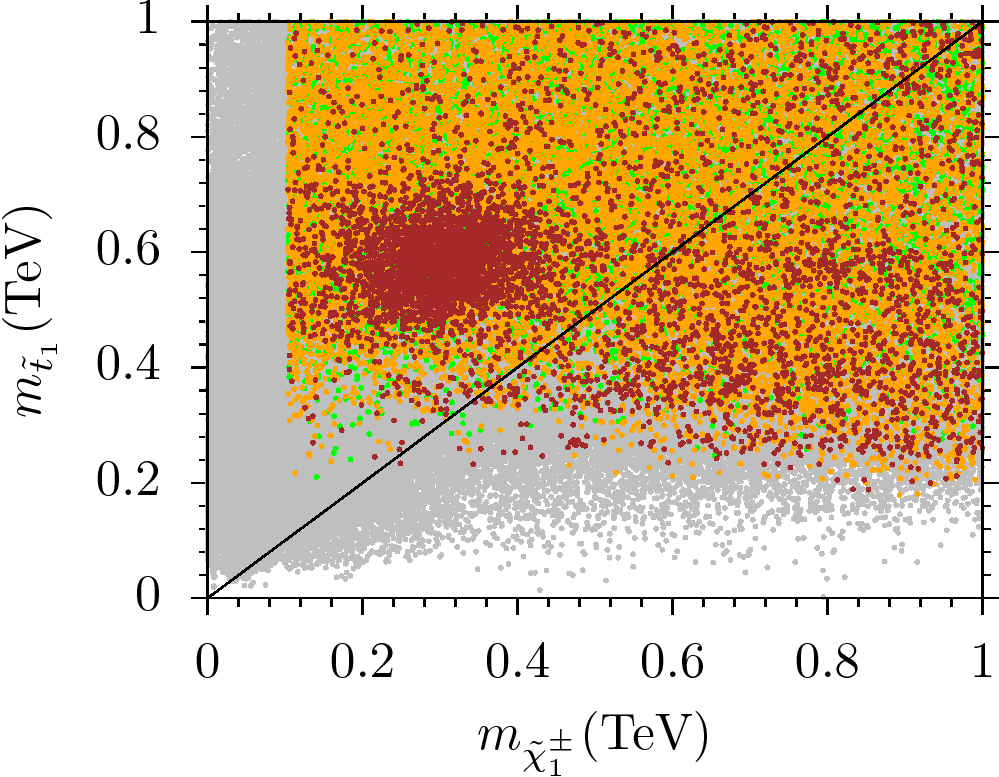}}
\caption{Plots in the $m_{\tilde{t}_{1}}-m_{\tilde{\chi}_{1}^{0}}$, $m_{\tilde{t}_{1}}-m_{\tilde{\chi}_{1}^{\pm}}$ planes. The color coding is the same as the right panel of Figure \ref{fig3}. The diagonal line indicates the mass degeneracy between the plotted particles.}
\label{fig5}
\end{figure}

Figure \ref{fig5} shows the masses of the lightest neutralino and the lightest chargino with plots in the $m_{\tilde{t}_{1}}-m_{\tilde{\chi}_{1}^{0}}$, $m_{\tilde{t}_{1}}-m_{\tilde{\chi}_{1}^{\pm}}$ planes. The color coding is the same as the right panel of Figure \ref{fig3}. The diagonal line indicates the mass degeneracy between the plotted particles. These two supersymmetric particles play a crucial role, since they take part in stop decay cascades, and the strictest constraints from the direct search at the LHC are based on the decay channels involving with the neutralino and chargino. Since we accept only the solutions yielding one of the neutralinos to be LSP, the final states of stop decays should include the neutralino. As mentioned before, the strongest bound on the stop mass is provided from the $\tilde{t}\rightarrow t \tilde{\chi}_{1}^{0}$ processes, and this decay channel is kinematically allowed only when $m_{\tilde{t}_{1}} \gtrsim m_{\tilde{\chi}_{1}^{0}}+m_{t}$. The $m_{\tilde{t}_{1}}-m_{\tilde{\chi}_{1}^{0}}$ plane shows that the LSP neutralino can be even almost massless, and hence the $\tilde{t}\rightarrow t \tilde{\chi}_{1}^{0}$ can be realized even when $m_{\tilde{t}_{1}} \sim 200$ GeV. A similar discussion can be followed when stop decays into a bottom quark and chargino, which bounds the stop mass as $m_{\tilde{t}}\gtrsim 650$ GeV. Indeed this channel is the best option to analyze and exclude the stop solutions below some scales. The lightest chargino mass is realized as low as about 100 GeV as seen from the $m_{\tilde{t}_{1}}-m_{\tilde{\chi}_{1}^{\pm}}$ plane. Since the mass of the bottom quark is negligible in compared to the stop and chargino masses, the $\tilde{t}\rightarrow b \tilde{\chi}_{1}^{0}$ can be realized even when the stop and chargino are nearly degenerate in mass.  

Before concluding, all the highlighted stop masses are excluded by the current LHC results. Despite the confidentiality of such strict constraints over the low scale analyses, some assumptions behind such experimental analyses may not be fulfilled when the parameter space is constrained from the GUT scale. For instance, even though the the best exclusion channel is $\tilde{t}\rightarrow b \tilde{\chi}_{1}^{0}$, the chargino in this process should eventually decay into the neutralino along with appropriate SM particles, and the strict exclusion arises when chargino decays into a $W-$ boson and LSP neutralino. The stop and chargino decays can be linked to each other easily in the low scale considerations, since a large set of low scale free parameters of MSSM allows such freedom. On the other hand, the SUSY GUTs have only a few free parameters, and these two processes cannot set freely, but they are calculated in certain correlations. Thus, it is possible to find some solutions in which ${\rm BR}(\tilde{t}\rightarrow b \tilde{\chi}_{1}^{0}) \sim 1$, the chargino may not kinematically allowed to decay into a $W-$boson and LSP neutralino. In such cases, the largest branching ratio can be found for the processes in which the chargino decay in to $u\bar{d}\tilde{\chi}_{1}^{0}$. Such processes cannot provide strict constraints due to large uncertainties in the QCD sector. 

\section{LHC Escape of Light Stops}
\label{sec:LHC}

In this section, we discuss the possibility of the light stop solutions to survive or being excluded over some benchmark points. We consider the processes in which stop decays into either a top quark and LSP neutralino, or a bottom quark and chargino, which are the main channels in exclusion analyses. The latter processes have a large impact in excluding the light stop solutions when the chargino is allowed to decay into $W^{\pm}$ along with the LSP neutralino. As discussed before, SUSY GUTs can yield solutions which react different in such exclusive analyses, since some low scale fundamental parameters, such as mixings, masses, couplings of supersymmetric particles relevant to the analyzed processes, are determined and constrained by a few  GUT scale fundamental parameters. 

To investigate the impact of the negative results from the direct searches we follow similar analyses represented in \cite{Chatrchyan:2013xna}. Generating events for the signals and relevant bacground processes are performed by using MadGraph \cite{Alwall:2011uj}. We, then transfer the generated event files to Delphes \cite{deFavereau:2013fsa} to employ the detector response. Finally, the results are plotted by using MadAnalysis \cite{Conte:2012fm}. We also apply some cuts to suppress the background, which are also employed in the analyses represented in  \cite{Chatrchyan:2013xna}. These cuts can be listed as follows:

\begin{itemize}
\item $E_{{\rm T}}^{{\rm miss}} > 100$ GeV, $M_{T} > 120$ GeV,

\item $P_{T} > 30$ GeV and $|\eta| < 2.4$ for jets,

\item $P_{T} > 30(25)$ GeV and $|\eta| < 1.422 (2.1)$ for electrons (muons),

\item $P_{T}^{{\rm total}} < {\rm min}(5~{\rm GeV},0.15P_{T}^{l})$,

\item $\Delta R(j,l) > 0.4$.
\end{itemize}
where $E_{T}^{{\rm miss}}$ is the missing transverse energy, while $M_{T}$ represents the transverse mass, and $P_{T}$ stands for the transverse momentum. 

Both signal processes end up with the final cases involving with a pair of each b quark, charged lepton, neutrino and LSP neutralino. The signal processes have to include LSP neutralinos, since R-parity is conserved. The relevant background has a final state of all these particles except neutralino. Only a pair of neutrinos contributes to the missing energy in the background process, also the neutralinos contribute in the signals; and hence the cut on $E_{T}^{{\rm miss}}$ is useful to suppress the background. In addition, since there are more particles in the final states of signals, the transverse mass is expected to be greater than that of the background process. The cuts on the transverse momentum $P_{T}$ makes possible to isolate the leptons and $\Delta R (j,l) > 0.4$ prevent them to overlap with the jets. Even though we do not apply a specific cut on $P_{T}$ of b quarks ($P_{T}^{(b)}$), $P_{T}^{(b)}$ is expected to be greater for the signals than the background. 

The challenge in detecting the stop is that it yields quite similar final state configuration to those involving with top quark. Thus, despite the discussion about the cuts above, suppressing the background results in also significant suppression in the signal as well. We will consider two signal processes separately next, and discuss the overall results for the stop detection in details. Our analyses are performed for the collisions with 13-14 TeV center of mass energy, and we set the luminosity to $19.5$ fb$^{-1}$.

\subsection{$\mathbf{\tilde{t}\rightarrow t \tilde{\chi}_{1}^{0}}$}

We first discuss the process $pp \rightarrow \tilde{t}\tilde{t}^{*}\rightarrow t\bar{t}\tilde{\chi}_{1}^{0}\tilde{\chi}_{1}^{0}\rightarrow b\bar{b}W^{\pm}W^{\mp}\tilde{\chi}_{1}^{0}\tilde{\chi}_{1}^{0}\rightarrow b\bar{b}l^{\pm}\bar{l}^{\mp}\nu_{l}\nu_{l}\tilde{\chi}_{1}^{0}\tilde{\chi}_{1}^{0}$ over some benchmark points within $200 \lesssim m_{\tilde{t}} \lesssim 700$ GeV given in Table \ref{bench1}. All points are consistent with the current experimental results. All masses are in GeV unit, while the cross-sections are given in pb. As seen, all points predict ${\rm BR}(\tilde{t}\rightarrow) t\tilde{\chi}_{1}^{0} \sim 1$ and hence $m_{\tilde{t}-m_{\tilde{\chi}_{1}^{0}}} \geq m_{t}$. The relevant background process is $pp \rightarrow t\bar{t}\rightarrow b\bar{b}W^{\pm}W^{\mp}\rightarrow b\bar{b}l^{\pm}\bar{l}^{\mp}\nu_{l}\nu_{l}$.

\begin{figure}[ht!]
\centering
\subfigure{\includegraphics[scale=0.25]{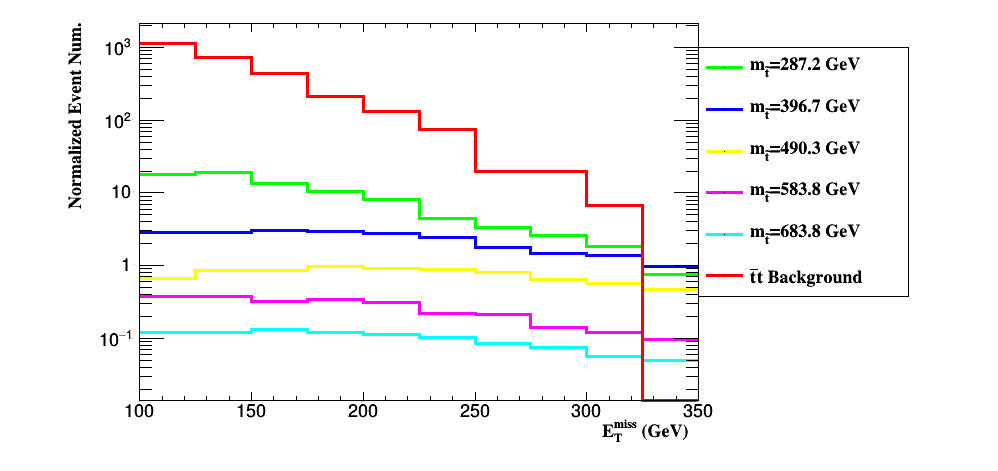}}%
\subfigure{\includegraphics[scale=0.25]{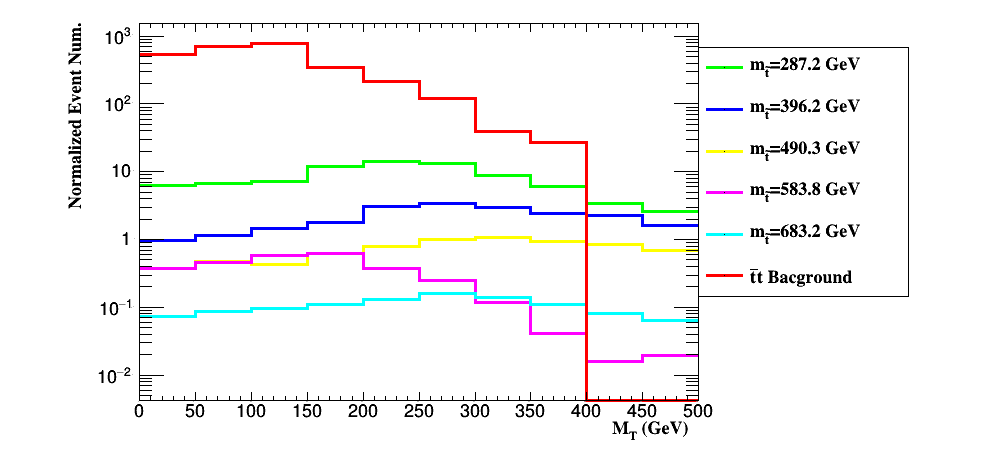}}
\caption{Plots representing the $E_{T}^{{\rm miss}}$ and $M_{T}$ for the signals and background. The cut on $E_{T}^{{\rm miss}}$ ($M_{T}$) is not applied on the left (right) panel.}
\label{fig:ETMT1}
\end{figure}

Figure \ref{fig:ETMT1} represents the $E_{T}^{{\rm miss}}$ and $M_{T}$ for the signals and background. The cut on $E_{T}^{{\rm miss}}$ ($M_{T}$) is not applied on the left (right) panel. As seen from the left panel, the cross-section for the most striking signal with $m_{\tilde{t}} \sim 290$ GeV is still about three magnitude smaller than the background, which leads to a small significance for the signal. As expected, the cross-section diminishes with the stop mass increasing. Even though the missing energy might be expected to be low for the background process, the energetic neutrinos can cause large missing energy, and it can be much larger than the cut applied on $E_{T}^{{\rm miss}}$. If one strengthens the cut on the missing energy as $E_{T}^{{\rm miss}} \lesssim 325$ GeV, then the background can be removed significantly. However,  the number of events in this region is $\lesssim 1$, and it is not enough for detection or exclusion at a high CL. Similar discussion can also be followed for the transverse mass as displayed in the right panel of Figure \ref{fig:ETMT1}. While a cut on $M_{T}$ applied as $M_{T} > 400$ GeV can remove the background, the signal processes cannot provide observable tracks, either. In this context, the cuts applied to suppress the background also suppress the signals significantly, and they result in quite a few number of events, which makes signals difficult to detect.  

The benchmark points considered as possible signals within $200 \lesssim m_{\tilde{t}} \lesssim 700$ GeV are listed in Table \ref{bench1}. All points are chosen as being consistent with the current experimental results. All masses are in GeV unit, while the cross-sections are given in pb. As seen, all points allow stop only to decay a top quark and LSP neutralino with the cross-section in a range as $10^{-3} \lesssim \sigma ({\rm signal}) \lesssim 10^{-1}$ pb. Although the benchmark points fulfill the assumption that is ${\rm BR}(\tilde{t}\rightarrow t\tilde{\chi}_{1}^{0}) \sim 1 $, they can be excluded only up to about $60\%$ CL for $m_{\tilde{t}} \lesssim 500$ GeV, while the exclusion cannot exceed a few percentage for $m_{\tilde{t}}\gtrsim 500$ GeV. 

\begin{table}[h!] 
\centering
\scalebox{0.7}{
\begin{tabular}{|c|ccccc|}
\hline
\hline
&&&&&\\
                & Point 1 & Point 2 & Point 3 & Point 4 & Point 5 \\
           &&&&&  \\
\hline
$m_{0}$       & 1773  & 2193  &  2551  & 2956 &  3164  \\
$M_{1} $      & -149.3  & -126.9  & -295.5 & -827.6 & -1006 \\
$M_{2} $      & -2848 & -3642 & -3904 & -6088  & -6290\\
$M_{3} $      & 795.8& 800.8 & 882.3& 1477 & 1519 \\
$\tan\beta$   &19.64 & 25.25 & 29.81 & 31.38 & 31.35\\
$A_0/m_{0}$   & -2.881 & -2.384 & -2.432  & -2.746 & -2.708  \\
$\mu$         & -1371 & -737.3 & -1123 & 799.1 & -1043 \\
$\Delta_{\rm EW}$   & 485.6 & 169.0 & 339.8  &  209.2 & 314.5\\
\hline
$m_h$         & 123.05 & 122.81 & 123.11  & 124.39 & 124.89 \\
$m_H$         & 2571 & 2750 & 2812 & 3625 & 3782 \\
$m_A$         & 2571 & 2750 & 2812 & 3625 & 3782 \\
$m_{H^{\pm}}$ & 2572 & 2751 & 2812 & 3626 & 3783 \\

\hline
$m_{\tilde{\chi}^0_{1,2}}$ & \textbf{ 66.76, 1380 } & \textbf{55.82, 743.2 } &\textbf{ 131.52, 1135  }  & \textbf{ 376.2, 808.9 }  &\textbf{459.2, 1055 } \\
$m_{\tilde{\chi}^0_{3,4}}$ &  1383, 2369 & 744.3, 3023 & 1136, 3247  &810.1, 5063 & 1056, 5186 \\
$m_{\tilde{\chi}^{\pm}_{1,2}}$ &\textbf{1380, 2369 }  & \textbf{ 742.3, 3024} & \textbf{ 1134, 3247 } & \textbf{ 807.9, 5063 }  & \textbf{ 1054, 5185 }\\
$m_{\tilde{g}}$  & 1919 & 1962 & 2150 &  3403 & 3498 \\
\hline $m_{ \tilde{u}_{L,R}}$ & 2938, 2322  & 3498, 2641 & 3875, 3006  & 5468, 3951 & 5664, 4145 \\
$m_{\tilde{t}_{1,2}}$ & \textbf{ 287.15, 2313 }  & \textbf{396.17, 2637 }  & \textbf{490.29, 3013}  & \textbf{ 583.81, 3960 }  &\textbf{ 683.82, 4162  }  \\
\hline $m_{ \tilde{d}_{L,R}}$ & 2938 , 2323 & 3498,2642  &  3875, 3006 &  5468, 3949 & 5664, 4142\\
$m_{\tilde{b}_{1,2}}$ & 2129, 2313 & 2334, 2637  & 2477, 3012  &  3216, 3957 & 3369, 4157 \\
\hline
$m_{\tilde{\nu}_{e,\mu}}$ &  2536, 2536& 3186, 3186 & 3548, 3547 & 4828 , 4827   &  5022, 5021\\
$m_{\tilde{\nu}_{\tau}}$ & 2461  &  3069 & 3357  &  4579  & 4759 \\
\hline
$m_{ \tilde{e}_{L,R}}$ & 2536, 1770   &  3186, 2189  & 3548, 2548  & 4828, 2965  & 5022, 3178   \\
$m_{\tilde{\tau}_{1,2}}$& 1570, 1770  &1863, 2189 & 1996, 2548 & 2101, 2965 & 2285, 3178 \\
\hline
$BR(\tilde{t}_{1} \rightarrow \tilde{\chi}^{0}_{1} t) $  & 1  & 1  & 1 &  1&  1 \\
$BR(\tilde{t}_{1} \rightarrow \tilde{\chi}^{\pm}_{1} b)$ & 0 & 0 & 0 &0 &0 \\
$BR({\tilde{{\chi}}}^{\pm}_{1} \rightarrow {\tilde{\chi}}^{0}_{1} W^{\pm}) $   & $ 3.4 \times 10^{-2}$ & $5.86 \times 10^{-2}$ & $4.6 \times 10^{-2}$ & $1.2 \times 10^{-1}$ & $8.7 \times 10^{-2}$ \\
\hline
 $\sigma({\rm signal})$ &$\mathbf{2.46\times 10^{-1} }$ & $\mathbf{4.544 \times 10^{-2}}$ & $\mathbf{1.375 \times 10^{-2}}$ &$\mathbf{4.883 \times 10^{-3}}$ &$\mathbf{1.855 \times 10^{-3}}$ \\
$\sigma(pp\rightarrow \tilde{t}\tilde{t}^{*})$ & $\mathbf{7.115 }$  & $\mathbf{1.314}$ & $\mathbf{ 4.018 \times 10^{-1}}$ & $\mathbf{ 1.545  \times 10^{-1}}$ & $\mathbf{5.523 \times 10^{-2}}$ \\
\hline
Exclusion CL\%  & 49.7 &   59.5 & 46.6 & 5.5 & 3.6  \\ 
\hline
\end{tabular}}
\caption{Benchmark points for $\tilde{t}\rightarrow t\tilde{\chi}_{1}^{0}$ within $200 \lesssim m_{\tilde{t}} \lesssim 700$ GeV. All points are chosen as being consistent with the current experimental results. All masses are in GeV unit, while the cross-sections are given in pb.}
\label{bench1} 
\end{table}
%%%%%%%%%%%%
%%%%%%%%%%
%%%%%%%%%%%%%%%%
%%%%%%%%%%%
%%%%%%%%%%%%%%%%%%%%%%%%%%%%%%%%%%%%%%%%%%

\subsection{$\mathbf{\tilde{t}\rightarrow b\tilde{\chi}_{1}^{\pm}}$}

We follow a similar analyses for the decay channel in which stop decays into a b quark and a chargino. The signal process can be expressed as $pp\rightarrow \tilde{t}\tilde{t}^{*}\rightarrow b\bar{b}\tilde{\chi}_{1}^{\pm}\tilde{\chi}_{1}^{\pm}\rightarrow b\bar{b}W^{\pm}W^{\pm}\tilde{\chi}_{1}^{0}\tilde{\chi}_{1}^{0}\rightarrow b\bar{b}l^{\pm}\bar{l}^{\pm}\nu_{l}\bar{\nu}_{l}\tilde{\chi}_{1}^{0}\tilde{\chi}_{1}^{0}$. In this case, it is not enough to have stop largely decay into a b quark and the chargino, since the chargino should also be allowed to decay into a $W-$boson and neutralino. If it is not allowed, the exclusion on the stop is not too much strict. The relevant background is the same as given for the previous signal processes. 

\begin{figure}[t!]
\centering
\subfigure{\includegraphics[scale=0.25]{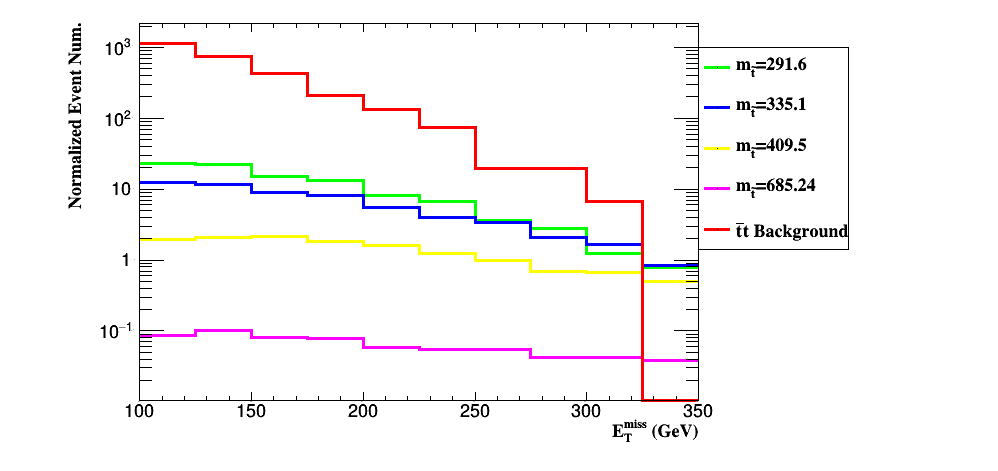}}%
\subfigure{\includegraphics[scale=0.25]{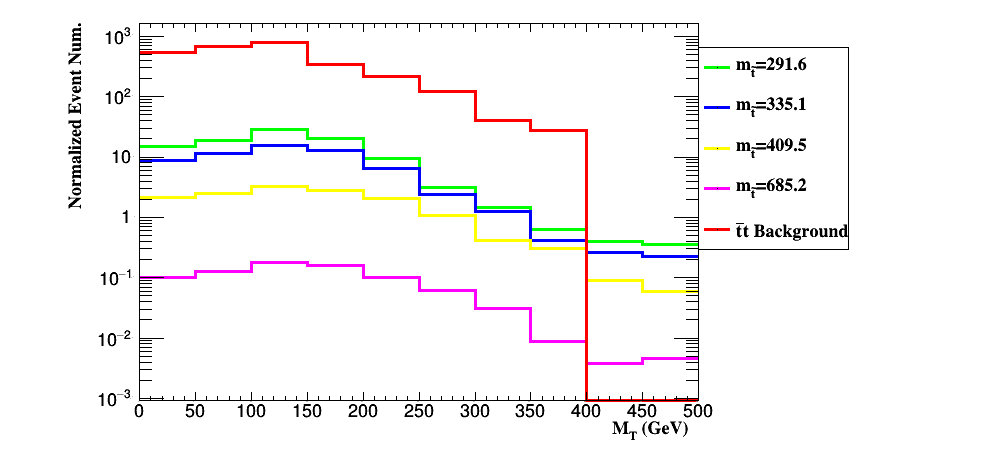}}
\caption{Plots representing the $E_{T}^{{\rm miss}}$ and $M_{T}$ for the signals and background. The cut on $E_{T}^{{\rm miss}}$ ($M_{T}$) is not applied on the left (right) panel.}
\label{fig:ETMT2}
\end{figure}

Figure \ref{fig:ETMT2} displays the plots for the $E_{T}^{{\rm miss}}$ and $M_{T}$ for the signals and background. The cut on $E_{T}^{{\rm miss}}$ ($M_{T}$) is not applied on the left (right) panel. A similar conclusion can be derived also for this type of signal processes, since suppressing the background does not leave enough number of events for the signal. In this context, the background mostly suppresses the signal processes and it is not possible to remove the background while keeping the signal processes intact. 

The details for the benchmark points considered as a possible signal are given in Table \ref{bench2} within $200 \lesssim m_{\tilde{t}} \lesssim 700$ GeV. All points are chosen as being consistent with the current experimental results. All masses are in GeV unit, while the cross-sections are given in pb. All points are chosen to yield the largest branching ratio for $\tilde{t}\rightarrow b\tilde{\chi}_{1}^{\pm}$. Note that we do not find any solution for $500 < m_{\tilde{t}} < 600$ GeV which allows large ${\rm BR}(\tilde{\chi}_{1}^{0}\rightarrow W^{\pm}\tilde{\chi}_{1}^{0})$. The cross-section for the signal changes from $\sim 10^{-1}$ pb to $\sim 10^{-3}$ pb. The exclusion can be as significant as about $45\%$ CL for $m_{\tilde{t}}\lesssim 500$ GeV, while excluding is not possible (by neglecting $1\%$ CL) for $m_{\tilde{t}} \gtrsim 500$ GeV. 

We consider the the decay channels which provides the strongest constraints on the stop mass, and we realized that the exclusion can be as good as only  at $50\% - 60\%$ CL. If one considers the $65\%$ CL as to be potentially observable signal and $95\%$ CL to be pure exclusion limit, the solutions with lighter stop mass can still have a chance to survive under the current collider analyses. Note that we set the luminosity close to its current values reached to the experiments. The number of events for the signals and consequently their exclusion level will raise with the increasing luminosity, and the exclusion will be severer near future.

 \begin{table}[t!] 
\centering
\scalebox{0.7}{
\begin{tabular}{|c|cccc|}
\hline
\hline
&&&&\\
                & Point 1 & Point 2 & Point 3 & Point 4 \\
           &&&& \\
\hline
$m_{0}$       & 1755 & 2062 &  2286 &  2925  \\
$M_{1} $      &   -151.8 & -182.8  & -135.2  & -1206   \\
$M_{2} $      & -3142 & -3799 & -3732 &	 -6053  \\
$M_{3} $      & 667.4 & 827.7 & 754.9 &  1467  \\
$\tan\beta$   & 22.6 & 22.4 & 23.7 &  31.54  \\
$A_0/m_{0}$   &  -2.33 & -2.38 &  -2.20 & -2.723  \\
$\mu$         & -189.0 & -216.9 & -236.7 & -641.8 \\
$\Delta_{\rm EW}$  & 40.7 & 56.7 & 52.8 & 139.4 \\
\hline
$m_h$         & 123.0 & 124.2 & 124.2  & 124.21 \\
$m_H$         & 2350 & 2808 & 2872 & 3580 \\
$m_A$         & 2350 & 2808 & 2872 & 3580 \\
$m_{H^{\pm}}$ & 2351 & 2808 & 2872 & 3580 \\

\hline
$m_{\tilde{\chi}^0_{1,2}}$ & \textbf{61.0, 179.5} & \textbf{75.7, 202.9} &\textbf{56.2, 226.2}  & \textbf{ 541.1, 651.1 }   \\
$m_{\tilde{\chi}^0_{3,4}}$ & 184.6, 2599 & 208.2, 3150 & 228.7, 3098 &  660.4, 5032 \\
$m_{\tilde{\chi}^{\pm}_{1,2}}$ &\textbf{173.7, 259.9}  & \textbf{ 198.0, 3150} & \textbf{221.2, 3098} & \textbf{649.7, 5032} \\
$m_{\tilde{g}}$  & 1656 & 2014 & 1871 &  3382 \\
\hline $m_{ \tilde{u}_{L,R}}$ & 2938, 2159 & 1113, 2573 & 1124, 2268 &  5417, 3912 \\
$m_{\tilde{t}_{1,2}}$ & \textbf{291.6, 2149}  & \textbf{335.1, 2558}  & \textbf{409.5, 2666}  & \textbf{685.24, 3934 }  \\
\hline $m_{ \tilde{d}_{L,R}}$ & 2938, 2160 & 1113, 2575 & 1124, 2669 & 5417, 3906 \\
$m_{\tilde{b}_{1,2}}$ & 1965, 2149 & 2344, 2558 & 2419, 2665 & 3194, 3927\\
\hline
$m_{\tilde{\nu}_{e,\mu}}$ & 2667, 2666 & 1005, 1005 & 1041, 1041 & 4790, 4789 \\
$m_{\tilde{\nu}_{\tau}}$ & 2589 & 3086 & 3191 & 4550\\
\hline
$m_{ \tilde{e}_{L,R}}$ & 2667, 1751 & 1005, 2058 & 1041, 2282 & 4790, 2951 \\
$m_{\tilde{\tau}_{1,2}}$& 1548, 1752 & 1818, 2058 & 2014, 2282 & 2105, 2951 \\
\hline
$BR(\tilde{t}_{1} \rightarrow \tilde{\chi}^{0}_{1} t) $  & 0.08 &  0.11 & 0.13& 0 \\
$BR(\tilde{t}_{1} \rightarrow \tilde{\chi}^{\pm}_{1} b)$ & 0.92&0.89 &0.7 & 1 \\
$BR({\tilde{{\chi}}}^{\pm}_{1} \rightarrow {\tilde{\chi}}^{0}_{1} W^{\pm}) $   & 1 & 1 & 1 & 1  \\
\hline
$\sigma(pp\rightarrow \tilde{t}\tilde{t}^{*})$ & $\mathbf{6.58}$  & $\mathbf{3.2}$ & $\mathbf{1.1}$ & $\mathbf{5.447 \times 10^{-2}}$ \\
$\sigma({\rm signal})$ &$\mathbf{2.4 \times 10^{-1}}$ & $\mathbf{1.1 \times 10^{-1}}$ & $\mathbf{ 2.0\times 10^{-2}}$ &$\mathbf{2.464\times 10^{-3}}$  \\
\hline
Exclusion  & $ 42.1\%$ CLs&  $ 42.8\%$ CLs& $ 43.3\%$ CLs & $ 1.0 \%$ CLs \\ 
\hline
\end{tabular}}
\caption{Benchmark points for $\tilde{t}\rightarrow b\tilde{\chi}_{1}^{\pm}$ within $200 \lesssim m_{\tilde{t}} \lesssim 700$ GeV. All points are chosen as being consistent with the current experimental results. All masses are in GeV unit, while the cross-sections are given in pb.}
\label{bench2} 
\end{table}

Before concluding this section, one also needs to discuss the reason why the cross-sections for the signals are at least three magnitude smaller than the background, despite the large branching ratios for the relevant decays of stops in the chosen benchmark points. The both signal processes start with the stop pair production, while the background include a pair of top quarks whose production cross-section is 

\begin{equation}
\sigma(pp\rightarrow t\bar{t})=\left[829\pm 50({\rm stat})\pm 56({\rm syst})\pm 83({\rm lumi}) \right]{\rm pb}
\label{topprod}
\end{equation}

Compared to the cross-section given in Eq.(\ref{topprod}), the largest cross-section for the stop pair productions are realized as  $\sim 7$ pb for the first points of Table \ref{bench1} and Table \ref{bench2}, which is much smaller even than errors in the production cross-section of the top quark pair. In this context, the stop pair production with a negligible cross-section is the main reason, which reduce the total cross-section in the considered signal processes.

\section{Conclusion}
\label{sec:conc}

We discussed the fine-tuning issue within the MSSM framework. We interpreted the fine-tuning as an indication for missing mechanisms, which can be left out in the minimal supersymmetric models. Following this idea we imposed non-universal gaugino masses at the GUT scale. We showed that the $\mu-$term is the main parameter which determines the required fine-tuning amount, and it is possible to realize $\mu \approx 0$ consistently with the EW breaking. Even though $\mu$ is the main parameter, it also has an impact on the SSB Higgs field mass, $m_{H_{u}}$, since $\mu \approx m_{H_{u}}$ is required to have the EW breaking at the correct scale ($M_{Z}\sim 90$ GeV). On the other hand, $m_{H_{d}}$ has almost no impact on the fine-tuning measurements, since its contributions are suppressed by $\tan\beta$. Any value of $\tan\beta$ can yield an acceptable amount of fine-tuning, but it is restricted to the range 10-30, if one also requires the solutions to yield light stop masses ($m_{\tilde{t}} \leq 700$ GeV). Even though we do not apply a direct bound on the stop mass, the other LHC constraints can bound the stop mass from below. The current results from the rare B-meson decays and the Higgs boson mass do not allow solutions with $m_{\tilde{t}} \lesssim 200$ GeV. In addition, the gluino mass bound ($m_{\tilde{g}} \geq 1.8$ TeV) excludes those with $m_{\tilde{t}} \lesssim 300$ GeV. However, it is still possible to realize $m_{\tilde{t}} \lesssim 400$ GeV, when $m_{\tilde{g}} \gtrsim 4$ TeV. We also observe that the mixing in the stop sector parametrized with $A_{t}$ has stronger impact than the stop mass itself on the required fine-tuning such that $\Delta _{EW}$ remains almost the same with constant $A_{t}$, even if the stop mass increases. In addition to the stop and gluino, sbottom mass lies from $600$ GeV to about 3 TeV, and it can be degenerate with stop when $m_{\tilde{t}}\sim m_{\tilde{b}} \approx 600$ GeV. Besides, the low scale mass spectrum yield $m_{\tilde{\tau}} \gtrsim 1$ TeV.

Finally we discussed the detection and exclusion possibility of the stops lighter than 700 GeV over the strict channels in the collider analyses. We chose benchmark points which predict the largest impact on the relevant decay channels within our data. We found that the largest cross-section obtained for a possible signal process is about three magnitude smaller than the background processes. The cuts applied to suppress the background left quite few number of events for the signals, which make the detection or exclusion obscure. The possible exclusion level is at about $60\%$CL at most for the processes involving with $\tilde{t}\rightarrow t \tilde{\chi}_{1}^{0}$, while it is about $50\%$ for those with $\tilde{t}\rightarrow b \tilde{\chi}_{1}^{\pm}$. These results are valid when $m_{\tilde{t}} \lesssim 500$ GeV, while the exclusion level significantly decreases when $m_{\tilde{t}} \gtrsim 500$ GeV. If one can require the exclusion at $65\%$ CL at least to have a clear signal, these exclusion levels are still lower, and the solutions with light stops may still have a chance to survive under the current limits. Note that the exclusion could be much severer, if the model was not constrained from the GUT scale. Despite large branching ratios predicted by the benchmark points, the small cross-sections for the signal processes arises from the stop pair production for which $\sigma(pp\rightarrow \tilde{t}\tilde{t}^{*}) \lesssim 7$ pb. Its cross-section is smaller even than the error bars in calculation of $\sigma(pp\rightarrow t\bar{t})$. Our analyses represented in this work were performed with 19.5 fb$^{-1}$ luminosity. The number of events for the signal processes will raise with the increasing luminosity; thus, one can conclude the the exclusion will be severer when the new results are released from near future experiments.

\noindent {\bf Acknowledgments}

We would like to thank Qaisar Shafi, Durmu\c{s} Ali Demir, Zafer Alt\i n, and B\"{u}\c{s}ra Ni\c{s} for fruitful discussions. We are also grateful to Jack Araz for useful discussions about LHC analyses. Part of the numerical calculations reported in this paper were performed at the National Academic Network and Information Center (ULAKBIM) of The Scientific and Technological Research Council of Turkey (TUBITAK), High Performance and Grid Computing Center (Truba Resources).

\end{document}